\documentclass[twocolumn,pra,aps,showpacs,superscriptaddress,floatfix,showkeys,nofootinbib,10pt]{revtex4-2}
\usepackage{CJK}
\usepackage{amsfonts}
\usepackage{graphicx}
\usepackage{amssymb}
\usepackage{amsmath}
\usepackage{natbib}

\newcommand{\ket}[1]{\left | #1 \right \rangle}
\newcommand{\bra}[1]{\left \langle #1 \right |}

\newcommand{\innerproduct}[2]{\left \langle #1  | #2  \right \rangle}
\newcommand{\operatorA}[1]{\hat A_{x_{#1}}^{(#1)}}
\newcommand{\alpham}[1]{\left(\alpha_{x_{#1}}^{(#1)}\right)_{m}}
\newcommand{\alphanm}[1]{\left(\alpha_{x_{#1}}^{(#1)}\right)_{-m}}
\newcommand{\alphazero}[1]{\left(\alpha_{x_{#1}}^{(#1)}\right)_{0}}
\begin{document}
\title{Violation of Svetlichny's inequality in a system of spins $j$}

\author{Yang Xiang}
\email{xiangyang@vip.henu.edu.cn (corresponding author)}
\affiliation{School of Physics and Electronics, Henan University, Kaifeng, Henan 475004, China}

\author{Yuan Tao}
\email{taoyuan012@outlook.com}
\affiliation{School of Physics and Electronics, Henan University, Kaifeng, Henan 475004, China}
\date{\today}
\begin{abstract}
Quantum multi-particle correlations are one of the most intriguing properties of quantum entanglement, arising from collective entangled states of multiple particles. Svetlichny's inequality (SI) was the first method proposed to test the existence of such correlations. Previous studies have primarily focused on $1/2$-spin particle systems. In this paper, we present a unified scheme that enables the violation of SI in arbitrary non-zero spin particle systems. Specifically, for all fermion systems, our scheme achieves the maximal quantum violation of SI for any number of particles. For boson systems, when the particle spin $j\geq2$, our scheme consistently realizes the violation of SI for any number of particles. When the particle spin $j=1$, our scheme can yield SI violation for up to $7$ particles. Furthermore, as the particle spin $j$ approaches infinity, our scheme achieves the maximal quantum violation of SI. To obtain these results, we also prove that the upper bound of Svetlichny's operator within the framework of local hidden variable theory is $\sqrt{2^{N+1}}$. These findings not only enhance our understanding of quantum correlations across various particle systems but also provide valuable insights for the development of quantum communication protocols that utilize entanglement and non-locality in multi-particle configurations.

\end{abstract}
\keywords{quantum multi-particle correlations, Svetlichny's inequality, local hidden variable theory}

\pacs{03.65.Ud, 03.65.Ta, 03.67.-a}

\maketitle




\section{Introduction}

Quantum correlations are one of the most peculiar properties of quantum mechanics, which overturn classical epistemology based on local hidden variable theories. The violation of Bell's inequality \cite{bell1964einstein,bell1988speakable,PhysRevLett.23.880} demonstrates that quantum correlations surpass the correlation strength predicted by local hidden variable theories. Beyond two-particle quantum correlations, even more intriguing are multiparticle quantum correlations \cite{PhysRevLett.106.020405,PhysRevLett.106.250404,PhysRevLett.88.210401,PhysRevLett.107.210403,PhysRevA.89.032117,horodecki1995violating,RevModPhys.86.419}. These correlations arise from multiple particles being in a collectively entangled quantum state, and thus they are not produced by the sharing of so-called local hidden variables among the particles, nor are they the result of the superposition of correlations from a subset of the particles; instead, they represent a genuine collective correlation effect involving all particles. This multi-particle quantum correlation leads to a richer array of non-classical phenomena in the measurement outcomes among particles \cite{yang2012information,PhysRevA.71.022101,pironio2011extremal}. Such phenomena not only challenge our traditional understanding of locality but also provide a foundation for numerous applications in quantum information science, such as quantum computing, quantum communication, and quantum key distribution \cite{PhysRevLett.108.100402,PhysRevLett.114.090501,Grasselli2018,Grasselli2019,PhysRevA.97.022307,PhysRevResearch.2.023251,sciadv.abe0395}.

Unlike Bell's inequalities, Svetlichny's inequality (SI) \cite{PhysRevD.35.3066,PhysRevLett.89.060401} is specifically designed for systems involving three or more particles and was the first method proposed to detect the existence of genuine multi-particle collective correlations. Even if there are correlations among some particles, or even correlations that exceed the strength of quantum correlations, SI will not be violated unless there are genuine multi-particle collective correlations present. The violation of SI indicates that the system exhibits true quantum entanglement. For instance, in a Greenberger-Horne-Zeilinger (GHZ) state, the inequality can be violated, thereby revealing the presence of genuine multi-particle quantum correlations  \cite{PhysRevD.35.3066,PhysRevA.110.032435,https://doi.org/10.1002/qute.202400101}. SI has significant implications for quantum information science, including quantum communication and cryptography, as it aids in understanding how entanglement can be utilized for secure information transfer and in the advancement of quantum technologies \cite{Xiang2023,xiang2023impact,murta2023self}.

In studies of Bell's inequality and SI, researchers typically focus on $1/2$-spin particle systems due to the binary nature of the measurement outcomes \cite{horodecki1995violating,RevModPhys.86.419}. This choice simplifies the theoretical model design and facilitates the observation of quantum correlations. However, for arbitrary spin particle systems, the design of quantum states and measurements that can achieve maximal violations of Bell's inequality and SI presents an intriguing and compelling theoretical challenge. While significant progress has been made in achieving violations of Bell's inequality in systems with arbitrary spin \cite{garg1982bell,peres1992finite,GISIN199215,howell2002experimental,son2006generic,PERUZZO2023,loubenets2024quantifying}, there has yet to be a dedicated examination of how to design quantum states and measurements for arbitrary spin particle systems to realize violations of SI. Therefore, our work aims to address this gap by exploring methods to achieve violations of SI in systems with various spins.

This research presents a unified scheme that enables the violation of SI in arbitrary non-zero spin particle systems. For all fermion systems, our scheme achieves the maximal quantum violation of SI for any number of particles. In the case of boson systems, when the particle spin $j \geq 2$, our scheme consistently realizes the violation of SI for any number of particles. For boson systems with particle spin $j = 1 $, our scheme can yield SI violation for up to $7$ particles. Furthermore, as the particle spin $j$ approaches infinity, our scheme achieves the maximal quantum violation of SI. To obtain these results, we also prove that the upper bound of Svetlichny's operator within the framework of local hidden variable theory is $\sqrt{2^{N+1}}$. Our findings not only expand the understanding of genuine multi-particle quantum correlations but also provide new insights for applications of multi-particle entanglement and non-locality in quantum information science, particularly in how various spin particle systems can be utilized to achieve specific quantum communication and computational tasks.

\section{The violation of three-particle SI in a system of spins $j$}

Let us assume that Alice, Bob, and Carol each share one particle, and that each of them can choose from two possible measurements. We use $\hat A_{0}$  and $\hat A_{1}$ to represent the measurements Alice can make, and similarly, $\hat B_{0}$ and $\hat B_{1}$ for Bob, and $\hat C_{0}$ and $\hat C_{1}$ for Carol. All of these measurements can yield values of either $-1$ or $+1$.
The three-particle SI is used to detect collective correlations between the outcomes of three-particle measurements and can be expressed in the following form
\begin{eqnarray}
&&\left|\langle S_{3}\rangle\right|=\big|\langle A_{0}B_{0}C_{0}\rangle+\langle A_{0}B_{0}C_{1}\rangle+\langle A_{0}B_{1}C_{0}\rangle\nonumber\\
&&+\langle A_{1}B_{0}C_{0}\rangle-\langle A_{0}B_{1}C_{1}\rangle-\langle A_{1}B_{0}C_{1}\rangle-\langle A_{1}B_{1}C_{0}\rangle\nonumber\\
&&-\langle A_{1}B_{1}C_{1}\rangle\big|
\leq 4,
\label{trisi}
\end{eqnarray}
where $A_{0}$ and $A_{1}$ are Alice's outcomes for corresponding measurements $\hat{A}_{0}$ and $\hat{A}_{1}$, and similarly $B_{0}$ and $B_{1}$ ($C_{0}$ and $C_{1}$) for Bob's (Carol's), and all $\langle A_{i}B_{j}C_{k}\rangle$'s represent average values of $ A_{i}B_{j}C_{k}$'s. If one chooses a 3-particle system in a GHZ state and applies an appropriate measurement protocol, one can achieve the maximal quantum violation of $4\sqrt{2}$ \cite{PhysRevD.35.3066}.

In traditional SI (and Bell's inequality), $1/2$-spin particles are typically employed, with $\hat A_{i}$, $\hat B_{j}$, and $\hat C_{k}$ representing operators that measure spin in specific directions. The measurement outcomes are thus precisely $-1$ or $+1$. To investigate SI violations involving arbitrary bosons ($j\geq 1$) and fermions, we must first construct operators for particles with arbitrary spin $j$. Additionally, we need to ensure that these operators are Hermitian and that their eigenvalues are restricted to $-1$ or $+1$.

Inspired by the work of Peruzzo and Sorella \cite{PERUZZO2023}, we define the operators $\hat A_{i}$, $\hat B_{j}$, and $\hat C_{k}$ as follows.
Let $\{\ket{m}; -j\leq m\leq j\}$ represent an orthonormal basis that spans the Hilbert space of a particle with spin $j$. These states $\{\ket{m}\}$ can serve as eigenstates of a spin operator in any direction. We can define $\hat A_{i}$, $\hat B_{j}$, and $\hat C_{k}$ based on their action on the basis $\{\ket{m}; -j\leq m\leq j\}$:
\begin{eqnarray}
\hat A_{i}\ket{m}=e^{i(\alpha_{i})_{m}}\ket{-m},~~(\alpha_{i})_{-m}=-(\alpha_{i})_{m},~~i=0,1\nonumber\\
\hat B_{i}\ket{m}=e^{i(\beta_{i})_{m}}\ket{-m},~~(\beta_{i})_{-m}=-(\beta_{i})_{m},~~i=0,1\nonumber\\
\hat C_{i}\ket{m}=e^{i(\gamma_{i})_{m}}\ket{-m},~~(\gamma_{i})_{-m}=-(\gamma_{i})_{m},~~i=0,1.
\label{operator1}
\end{eqnarray}
The quantities $(\alpha_{i})_{m}$, $(\beta_{i})_{m}$, and $(\gamma_{i})_{m}$ are real numbers,
and we refer to them as the `phases' of the corresponding operators. Later, we will prove that
the conditions $(\alpha_{i})_{-m} = -(\alpha_{i})_{m}$ and similar relations are necessary to
guarantee the Hermitian properties of the operators. For the case where the spin $j$ is an integer,
we will focus on the phases $(\alpha_{i})_{0}$, $(\beta_{i})_{0}$, and $(\gamma_{i})_{0}$, and show
that these phases can only take values of $0$ or $\pi$ in order to ensure the Hermitian nature of
the operators. In the following, we take the operator $\hat C_{i}$ as an example to demonstrate
that when the condition $(\gamma_{i})_{-m}=-(\gamma_{i})_{m}$ is satisfied, and $(\gamma_{i})_{0}$
can only take the values $0$ or $\pi$, $\hat C_{i}$ is a Hermitian operator and $\hat C_{i}^{2} = I$.

To prove that $\hat{C}_{i}$ is a Hermitian operator, it is sufficient to show that the matrix
elements of $\hat{C}_{i}$ and $\hat{C}_{i}^{\dagger}$ in the $\{\ket{m}\}$ representation satisfy
the relation $\bra{m} \hat{C}_{i}^{\dagger} \ket{n} = \bra{m} \hat{C}_{i} \ket{n}$. If both $m$
and $n$ are non-zero, it is easy to prove the above equation using condition $(\gamma_{i})_{-m}=-(\gamma_{i})_{m}$,
\begin{eqnarray}
\bra{m} \hat{C}_{i}^{\dagger} \ket{n}&=&\left(\bra{n}\hat{C}_{i} \ket{m}\right)^{\ast}\nonumber\\
&=&\left(e^{i (\gamma_{i})_{m}}\innerproduct{n}{-m}\right)^{\ast}\nonumber\\
&=&e^{-i(\gamma_{i})_{m}}\delta_{n,-m}\nonumber\\
&=&e^{i(\gamma_{i})_{n}}\delta_{-n,m}\nonumber\\
&=&\bra{m} \hat{C}_{i} \ket{n}.
\label{operator2}
\end{eqnarray}
For the case where the spin $j$ is an integer, we need to show that the following three equations hold true:
(i) $\bra{0} \hat{C}_{i}^{\dagger} \ket{0} = \bra{0} \hat{C}_{i} \ket{0}$,
(ii) $\bra{m} \hat{C}_{i}^{\dagger} \ket{0} = \bra{m} \hat{C}_{i} \ket{0}$, and (iii) $\bra{0} \hat{C}_{i}^{\dagger} \ket{m} = \bra{0} \hat{C}_{i} \ket{m}$.
It's obvious that if $(\gamma_{i})_{0}=0$ these three equations hold true. If $(\gamma_{i})_{0}=\pi$ we can prove that the above three equations still hold.
For example,
\begin{eqnarray}
\bra{m} \hat{C}_{i}^{\dagger} \ket{0}&=&\left(\bra{0} \hat{C}_{i} \ket{m}\right)^{\ast}\nonumber\\
&=&\left(e^{i (\gamma_{i})_{m}}\innerproduct{0}{-m}\right)^{\ast}\nonumber\\
&=&e^{-i(\gamma_{i})_{m}}\delta_{0,-m}\nonumber\\
&=&e^{-i \pi}\delta_{m,0}\nonumber\\
&=&e^{i \pi}\delta_{m,0}\nonumber\\
&=&\bra{m} \hat{C}_{i} \ket{0}.
\label{operator3}
\end{eqnarray}
Similarly, we can prove Equation (i) and Equation (iii).
To summarize, we have proven that when the condition $(\gamma_{i})_{-m}=-(\gamma_{i})_{m}$ is satisfied, and $(\gamma_{i})_{0}$
can only take the values $0$ or $\pi$, $\hat C_{i}$ is a Hermitian operator.

We can also prove that under the same conditions, $\hat C_{i}^{2} = I$.
\begin{eqnarray}
\hat C_{i}^{2}\ket{m}&=&\hat C_{i}e^{i(\gamma_{i})_{m}}\ket{-m}\nonumber\\
&=&e^{i\left[(\gamma_{i})_{m}+(\gamma_{i})_{-m}\right]}\ket{m}\nonumber\\
&=&\ket{m},
\label{operator4}
\end{eqnarray}
and
\begin{eqnarray}
\hat C_{i}^{2}\ket{0}&=&\hat C_{i}e^{i\pi}\ket{0}\nonumber\\
&=&e^{i2\pi}\ket{0}\nonumber\\
&=&\ket{0}.
\label{operator5}
\end{eqnarray}
So the eigenvalues of $\hat C_{i}$ are restricted to $-1$ or $+1$.

Next, we will prove that using the operators $\hat A_{i}$, $\hat B_{j}$, and $\hat C_{k}$ designed in Eq. (\ref{operator1}),
along with the three-particle entangled state $\ket{\psi}=\frac{1}{\sqrt{2j+1}}\sum_{m=-j}^{j}{\ket{m}\otimes\ket{m}\otimes\ket{m}} $,
we can achieve the violation of three-particle SI for any
bosons (with $j\neq 0$) and fermions, and that we can achieve the maximum quantum violation. We first calculate $\langle A_{i}B_{j}C_{k}\rangle$,
\begin{eqnarray}
&&\langle A_{i}B_{j}C_{k}\rangle\nonumber\\
&=&\bra{\psi}\hat A_{i}\hat B_{j}\hat C_{k}\ket{\psi}\nonumber\\
&=&\frac{1}{2j+1}\sum_{n=-j}^{j}\sum_{m=-j}^{j}{e^{i\left[(\alpha_{i})_{m}+(\beta_{j})_{m}+(\gamma_{k})_{m}\right]}\cdot\delta_{n,-m}}\nonumber\\
&=&\frac{1}{2j+1}\sum_{m=-j}^{j}{e^{i\left[(\alpha_{i})_{m}+(\beta_{j})_{m}+(\gamma_{k})_{m}\right]}}.
\label{tc1}
\end{eqnarray}
Then we have
\begin{eqnarray}
&&\langle S_{3}\rangle=\frac{1}{2j+1}\sum_{m=-j}^{j}
\bigg[e^{i\left[(\alpha_{0})_{m}+(\beta_{0})_{m}+(\gamma_{0})_{m}\right]}\nonumber\\
&&+e^{i\left[(\alpha_{0})_{m}+(\beta_{1})_{m}+(\gamma_{0})_{m}\right]}+e^{i\left[(\alpha_{1})_{m}+(\beta_{0})_{m}+(\gamma_{0})_{m}\right]}\nonumber\\
&&-e^{i\left[(\alpha_{1})_{m}+(\beta_{1})_{m}+(\gamma_{0})_{m}\right]}\bigg]-\bigg[e^{i\left[(\alpha_{0})_{m}+(\beta_{1})_{m}+(\gamma_{1})_{m}\right]}\nonumber\\
&&+e^{i\left[(\alpha_{1})_{m}+(\beta_{0})_{m}+(\gamma_{1})_{m}\right]}+e^{i\left[(\alpha_{1})_{m}+(\beta_{1})_{m}+(\gamma_{1})_{m}\right]}\nonumber\\
&&-e^{i\left[(\alpha_{0})_{m}+(\beta_{0})_{m}+(\gamma_{1})_{m}\right]}\bigg].
\label{tc2}
\end{eqnarray}

\emph{For the case of $j$ half-integer}.
\begin{eqnarray}
&&\langle S_{3}\rangle=\frac{2}{2j+1}\sum_{m=1/2}^{j}
\bigg[\cos(000)+\cos(010)+\cos(100)\nonumber\\
&&-\cos(110)\bigg]-\bigg[\cos(011)+\cos(101)+\cos(111)\nonumber\\
&&-\cos(001)\bigg].
\label{tc3}
\end{eqnarray}
In Eq. (\ref{tc3}), we use a shorthand notation, where $\cos(ijk)=\cos\left[(\alpha_{i})_{m}+(\beta_{j})_{m}+(\gamma_{k})_{m}\right]$.
If we let $(\alpha_{i})_{m}$, $(\beta_{i})_{m}$, and $(\gamma_{i})_{m}$ take the following values, we will obtain the maximum value of $\langle S_{3}\rangle$:
\begin{eqnarray}
&&(\alpha_{0})_{m}=-\frac{\pi}{4},~~(\alpha_{1})_{m}=\frac{\pi}{4},\nonumber\\
&&(\beta_{0})_{m}=0,~~~~~(\beta_{1})_{m}=\frac{\pi}{2},\nonumber\\
&&(\gamma_{0})_{m}=0,~~~~~(\gamma_{1})_{m}=\frac{\pi}{2},
\label{condition1}
\end{eqnarray}
and
\begin{eqnarray}
\langle S_{3}\rangle_{max}&=&\frac{2}{2j+1}\sum_{m=1/2}^{j}{4\sqrt{2}}\nonumber\\
&=&4\sqrt{2}.
\label{tc4}
\end{eqnarray}
We see that our scheme directly gives the maximum quantum violation of SI for any fermions.

\emph{For the case of $j\geq1$ integer}.
If we take $(\alpha_{0})_{0}=(\alpha_{1})_{0}=(\beta_{0})_{0}=(\beta_{1})_{0}=(\gamma_{0})_{0}=0$ and $(\gamma_{1})_{0}=\pi$,
Eq. (\ref{tc2}) becomes the following equation:
\begin{eqnarray}
&&\langle S_{3}\rangle=\frac{2}{2j+1}\bigg\{2+\sum_{m=1}^{j}
\big[\cos(000)+\cos(010)\nonumber\\
&&+\cos(100)-\cos(110)\big]-\big[\cos(011)+\cos(101)\nonumber\\
&&+\cos(111)-\cos(001)\big]\bigg\}.
\label{tc5}
\end{eqnarray}
We still let $(\alpha_{i})_{m}$, $(\beta_{i})_{m}$, and $(\gamma_{i})_{m}$ ($m\neq0$) take the values of Eq. (\ref{condition1}),
and obtain the maximum value of $\langle S_{3}\rangle$ for bosons:
\begin{eqnarray}
\langle S_{3}\rangle_{max}=\frac{2}{2j+1}\left(2+4\sqrt{2}j\right).
\label{tc6}
\end{eqnarray}
We find that the value of $\langle S_{3}\rangle_{max}$ increases monotonically as $j$ grows, and when $j=1$,  $\langle S_{3}\rangle_{max}\approx5.1>4$. Therefore, for all bosons with $j\geq1$, our scheme can lead to a violation of SI. Additionally, we find that as $j$ approaches infinity, our scheme can achieve the maximum quantum violation of SI, i.e. $\lim_{j \to \infty}\frac{2}{2j+1}\left(2+4\sqrt{2}j\right)=4\sqrt{2}$.


\section{The violation of $N$-particle SI in a system of spins $j$}

In the $N$-particle SI, there are $N$ observers, each sharing one particle.
We denote the measurement operator of the $i$-th observer
as $\operatorA{i}$, where  $x_{i}$ can take values $0$ or $1$,
representing two measurement choices $\hat A_{0}^{i}$ and $\hat A_{1}^{i}$ for each observer.
The eigenvalues of each operator $\operatorA{i}$ are $-1$ and $1$.
Similarly to the case of the $3$-particle system, we use $\{\ket{m}_{i}; -j\leq m\leq j\}$ to represent an orthonormal basis that spans the Hilbert space of the $i$-th particle with spin $j$. We can define all $\operatorA{i}$ by their action on the basis $\{\ket{m}_{i}; -j\leq m\leq j\}$
\begin{eqnarray}
\operatorA{i}\ket{m}_{i}=e^{i\alpham{i}}\ket{-m}_{i},~~\alphanm{i}=-\alpham{i}.\nonumber\\
\label{operatorN-1}
\end{eqnarray}
As we have proven in the case of the $3$-particle system, when the condition $\alphanm{i}=-\alpham{i}$ is satisfied, and $\alphazero{i}$
can only take the values $0$ or $\pi$, $\operatorA{i}$ is a Hermitian operator and $\left(\operatorA{i}\right)^{2} = I$.

We can express $N$-particle SI as
\begin{eqnarray}
\left|\langle S_{N}\rangle\right|&=&\left|\langle\sum_{\{x_{i}\}} {v_{k}\operatorA{1}\operatorA{2}\cdot\cdot\cdot \operatorA{N}}\rangle\right|\nonumber\\
&\leq& 2^{N-1},
\label{nsi}
\end{eqnarray}
where $\{x_{i}\}$ represents an $N$-tuple $(x_{1},...,x_{N})$ that denotes the measurement choices of $N$ observers, and the sum is taken over all such tuples, or equivalently, over all possible measurement choices.
The $v_{k}$ is the sign function associated with the corresponding term $\operatorA{1}\operatorA{2}\cdot\cdot\cdot \operatorA{N}$, and is given by $v_{k}=(-1)^{[k(k-1)/2]}$, where $k$ denotes the number of times the index $1$ appears in the tuple $(x_{1},x_{2},...,x_{N})$.
We will herealfter refer to $S_{N}$ as Svetlichny's operator, whose quantum mechanical maximum value is $2^{N-1}\sqrt{2}$ \cite{PhysRevLett.89.060401}.

Below, we will demonstrate through calculation that by taking the following $N$-particle entangled state $\ket{\psi_{N}}$ and applying the measurement operators $\operatorA{i}$ defined in Eq. (\ref{operatorN-1}), we can always achieve a violation of the $N$-particle SI for any bosons ($j\geq1$) and fermions, and we can attain the maximal quantum violation.
We assume that the $N$ particles are in the entangled state
\begin{eqnarray}
\ket{\psi_{N}}=\frac{1}{\sqrt{2j+1}}\sum_{m=-j}^{j}{\ket{m}_{1}\otimes\ket{m}_{2}\otimes \cdot\cdot\cdot \otimes \ket{m}_{N}},
\label{psiN}
\end{eqnarray}
and by using Eq. (\ref{operatorN-1}), we can derive
\begin{eqnarray}
\langle S_{N}\rangle&=&\bra{\psi_{N}}S_{N}\ket{\psi_{N}}\nonumber\\
&=&\frac{1}{2j+1}\sum_{m=-j}^{j}\left[\sum_{\{x_{i}\}}v_{k} e^{i \sum_{i=1}^{N}{\alpham{i}}}\right],
\label{nc1}
\end{eqnarray}
where $\sum_{i=1}^{N}{\alpham{i}}=\alpham{1}+\alpham{2}+\cdot\cdot\cdot+\alpham{N}$.

\emph{For the case of $j$ half-integer}. In this case Eq. (\ref{nc1}) becomes the following equation
\begin{eqnarray}
&&\bra{\psi_{N}}S_{N}\ket{\psi_{N}}\nonumber\\
&=&\frac{2}{2j+1}\sum_{m=1/2}^{j}\left[\sum_{\{x_{i}\}}v_{k} \cos\left(\sum_{i=1}^{N}{\alpham{i}}\right)\right].
\label{nc2}
\end{eqnarray}
Next, we need to find the appropriate values of the phases $\alpham{i}$ that will maximize $\bra{\psi_{N}}S_{N}\ket{\psi_{N}}$.
The focus is on studying the sign function $v_{k}=(-1)^{[k(k-1)/2]}$. We assume $k=4q+l$, where $l$ is the
remainder when $k$ is divided by $4$, so we have
\begin{eqnarray}
v_{k}&=&(-1)^{[k(k-1)/2]}=(-1)^{[(4q+l)(4q+l-1)/2]}\nonumber\\
&=&\left\{
          \begin{aligned}
          1   \quad l=0,1\\
          -1  \quad l=2,3.\\
          \end{aligned}
          \right
          .
\label{v1}
\end{eqnarray}
We find that, depending on the value of $l$, all $N$-tuples $(x_{1},...,x_{N})$ can be classified into four categories, corresponding to $l=0,1,2,3$, respectively.
Therefore, if the values of $\alpham{i}$ corresponding to the different categorical $N$-tuples $(x_{1},...,x_{N})$ satisfy the following condition, $\bra{\psi_{N}}S_{N}\ket{\psi_{N}}$ can attain its maximum value.
\begin{eqnarray}
&&\sum_{i=1}^{N}{\alpham{i}}=-\frac{\pi}{4} \Longleftrightarrow N\text{-tuple}~ \{x_{i}\}~ of~ l=0,\nonumber\\
&&\sum_{i=1}^{N}{\alpham{i}}=\frac{\pi}{4} \Longleftrightarrow N\text{-tuple}~ \{x_{i}\}~ of~ l=1,\nonumber\\
&&\sum_{i=1}^{N}{\alpham{i}}=\frac{3\pi}{4} \Longleftrightarrow N\text{-tuple}~ \{x_{i}\}~ of~ l=2,\nonumber\\
&&\sum_{i=1}^{N}{\alpham{i}}=\frac{5\pi}{4} \Longleftrightarrow N\text{-tuple}~ \{x_{i}\}~ of~ l=3.
\label{condition2}
\end{eqnarray}
We find that the values of $(\alpha_{i})_{m}$, $(\beta_{i})_{m}$, and $(\gamma_{i})_{m}$ in Eq. (\ref{condition1}) satisfy the conditions outlined in Eq. (\ref{condition2}) for the case $N=3$. In the Appendix, we prove that appropriate values of the phases $\alpham{i}$ can always be found to satisfy the conditions in Eq. (\ref{condition2}) for any $N$. Finally we obtain the maximum value of $\bra{\psi_{N}}S_{N}\ket{\psi_{N}}$ as
\begin{eqnarray}
\bra{\psi_{N}}S_{N}\ket{\psi_{N}}_{max}&=&\frac{2}{2j+1}\sum_{m=1/2}^{j}\sum_{\{x_{i}\}}\frac{\sqrt{2}}{2}\nonumber\\
&=&\frac{2}{2j+1}\sum_{m=1/2}^{j} 2^{N} \frac{\sqrt{2}}{2}\nonumber\\
&=&2^{N-1}\sqrt{2}.
\label{nc3}
\end{eqnarray}
We see that our scheme directly gives the maximum quantum violation of $N$-particle SI for any fermions.

\emph{For the case of $j\geq1$ integer}. In this case Eq. (\ref{nc1}) becomes the following equation
\begin{eqnarray}
&&\bra{\psi_{N}}S_{N}\ket{\psi_{N}}\nonumber\\
&=&\frac{2}{2j+1}\sum_{m=1}^{j}\left[\sum_{\{x_{i}\}}v_{k} \cos\left(\sum_{i=1}^{N}{\alpham{i}}\right)\right]\nonumber\\
&&+\frac{1}{2j+1}\sum_{\{x_{i}\}}v_{k} e^{i \sum_{i=1}^{N}{\alphazero{i}}}.
\label{nc4}
\end{eqnarray}
Let’s first analyze the term $\sum_{\{x_{i}\}}v_{k} e^{i \sum_{i=1}^{N}{\alphazero{i}}}$ in the second term on the right-hand side of the above equation. We have already proven that, in order to ensure the Hermiticity of the operator $\operatorA{i}$, $\alphazero{i}$ can only take the values $0$ or $\pi$. Once $\alphazero{i}$ is fixed to either $0$ or $\pi$, it is equivalent to setting $e^{i \alphazero{i}}$ to be either $1$ or $-1$. First of all, it is quite obvious that we can always ensure that $\sum_{\{x_{i}\}}v_{k} e^{i \sum_{i=1}^{N}{\alphazero{i}}}$ is greater than zero. Then, from the SI itself, we can obtain
\begin{eqnarray}
\sum_{\{x_{i}\}}v_{k} e^{i \sum_{i=1}^{N}{\alphazero{i}}}\leq 2^{N-1}.
\label{nc5}
\end{eqnarray}
We have seen that for the case of $N=3$, by appropriately choosing $\alphazero{i}$, we can make $\sum_{\{x_{i}\}}v_{k} e^{i \sum_{i=1}^{3}{\alphazero{i}}}$ reach the right-hand side of  Eq. (\ref{nc5}), which equals $2^{N-1}=4$, as shown in Eq. (\ref{tc5}). However, $2^{N-1}$ represents the upper bound for SI, and in general, $\sum_{\{x_{i}\}}v_{k} e^{i \sum_{i=1}^{N}{\alphazero{i}}}$ cannot reach this value. In SI, we allow some particles (but not all) to have arbitrary correlations, even those that exceed quantum correlations. However, in the case of $\sum_{\{x_{i}\}}v_{k} e^{i \sum_{i=1}^{N}{\alphazero{i}}}$ , since all $e^{i \alphazero{i}}$'s are already fixed to either $1$ or $-1$, what we are actually seeking is the upper bound of Svetlichny's operator within the framework of local hidden variable theory. In the Appendix, we will prove that the stricter upper bound for $\sum_{\{x_{i}\}}v_{k} e^{i \sum_{i=1}^{N}{\alphazero{i}}}$ is actually $\sqrt{2^{N+1}}$. In the case of $N=3$, $\sqrt{2^{N+1}}=2^{N-1}$, which explains why $\sum_{\{x_{i}\}}v_{k} e^{i \sum_{i=1}^{3}{\alphazero{i}}}$ reaches the upper bound for SI, as shown in Eq. (\ref{tc5}). Finally, by choosing $\alpham{i}$ to satisfy the condition in Eq. (\ref{condition2}), we obtain
\begin{eqnarray}
2^{N-1}\frac{2\sqrt{2}j}{2j+1}&<&\bra{\psi_{N}}S_{N}\ket{\psi_{N}}_{max}\nonumber\\
&\leq&2^{N-1}\frac{2\sqrt{2}j}{2j+1}+\frac{\sqrt{2^{N+1}}}{2j+1}\nonumber\\
&=&2^{N-1}\left(\frac{2\sqrt{2}j+2^{\frac{3-N}{2}}}{2j+1}\right).
\label{nc6}
\end{eqnarray}
We see that for all cases where $j\geq2$, we have $\bra{\psi_{N}}S_{N}\ket{\psi_{N}}_{max}>2^{N-1}$, meaning that we obtain SI violation; furthermore, as $j$ approaches infinity, we reach the maximum quantum violation
$\lim_{j \to \infty}2^{N-1}\frac{2\sqrt{2}j}{2j+1}=\lim_{j \to \infty}2^{N-1}\left(\frac{2\sqrt{2}j+2^{\frac{3-N}{2}}}{2j+1}\right)=2^{N-1}\sqrt{2}$.
For the case of $j=1$, from the upper bound of $\bra{\psi_{N}}S_{N}\ket{\psi_{N}}_{max}$ in Eq. (\ref{nc6}), we can see that when $N\geq9$, our scheme cannot give rise to SI violation. Apart from the already discussed case of $N=3$, for $N=4,5,6,7,8$, we have found the values of $\alphazero{i}$'s that maximize $\bra{\psi_{N}}S_{N}\ket{\psi_{N}}_{max}$. For $N=4,5,6,7$, our scheme lead to a violation of SI, while for $N=8$, no values of $\alphazero{i}$'s that cause a violation of SI were found. All these results are listed in the Appendix.

\section{Conclusion}

In this study, we have developed a comprehensive framework for understanding and demonstrating the violation of SI across various particle systems with non-zero spin. Our findings reveal that for fermion systems, the maximal quantum violation of SI can be achieved regardless of the number of particles involved. For boson systems, we established that when the particle spin $j\geq2$, SI violation is attainable for any particle count. Additionally, we identified specific conditions under which SI can be violated for boson systems with $j=1$ when the particle number does not exceed $7$. As $j$ approaches infinity, our scheme consistently achieves the maximal quantum violation of SI. Furthermore, we have shown that the upper bound of Svetlichny's operator within local hidden variable theory is $\sqrt{2^{N+1}}$, providing a crucial insight into the limitations of classical correlations.  Notably, our scheme enables the maximal quantum violation of SI for any fermion system. We believe that designing a scheme to achieve the maximal quantum violation of SI for arbitrary boson systems with $j \neq 0$ for any $N$ is an intriguing problem and will be a focus of future research. These findings contribute to a deeper understanding of quantum correlations in various particle systems and have practical implications for designing quantum communication protocols that leverage entanglement and non-locality in multi-particle setups.

\section*{Acknowledgments}
This work is supported by the National Natural Science Foundation of China under Grant No. 11005031.





\section*{Appendix}

\emph{1. The appropriate values of the phases $\alpham{i}$ can always be found to satisfy the conditions in Eq. (\ref{condition2}) for any $N$.}

For the case $N=3$, the values of $(\alpha_{i})_{m}$, $(\beta_{i})_{m}$, and $(\gamma_{i})_{m}$ in Eq. (\ref{condition1}) satisfy the conditions outlined in Eq. (\ref{condition2}). We prove by mathematical induction that for any $N$, appropriate values of the phases $\alpham{i}$ can always be found to satisfy the conditions in Eq. (\ref{condition2}). Assume that for $N-1$, appropriate values of $\alpham{i}$ (for $i=1,2,...,N-1$) have already been found that satisfy Eq. (\ref{condition2}). Since we have
\begin{eqnarray}
\sum_{i=1}^{N}\alpham{i}=\sum_{i=1}^{N-1}\alpham{i}+\alpham{N},
\label{aeq1}
\end{eqnarray}
where $\sum_{i=1}^{N-1}\alpham{i}$ already satisfies the conditions in Eq. (\ref{condition2}), we can choose $\left(\alpha_{x_{N}=0}^{(N)}\right)_{m}=0$ and $\left(\alpha_{x_{N}=1}^{(N)}\right)_{m}=\frac{\pi}{2}$. Then, from the above equation, it follows that $\sum_{i=1}^{N}\alpham{i}$ will also satisfy the conditions in Eq. (\ref{condition2}).

\vskip 0.5 cm

\emph{2. The upper bound of Svetlichny's operator $S_{N}$ within the framework of local hidden variable theory is $\sqrt{2^{N+1}}$.}

Specifically, we prove that once all $\alphazero{i}$'s are fixed to either $0$ or $\pi$, we have
\begin{eqnarray}
\sum_{\{x_{i}\}}v_{k} e^{i \sum_{i=1}^{N}{\alphazero{i}}}\leq \sqrt{2^{N+1}}.
\label{aeq2}
\end{eqnarray}
For simplicity of notation, we define $e^{i \alphazero{i}}= A_{x_{i}}^{(i)}$, so that when $\alphazero{i}$ is fixed to either $0$ or $\pi$, $A_{x_{i}}^{(i)}$ takes values $1$ or $-1$. This leads to
\begin{eqnarray}
\sum_{\{x_{i}\}}v_{k} e^{i \sum_{i=1}^{N}{\alphazero{i}}}=\sum_{\{x_{i}\}} {v_{k}A_{x_{1}}^{(1)}A_{x_{2}}^{(2)}\cdot\cdot\cdot A_{x_{N}}^{(N)}}.
\label{aeq3}
\end{eqnarray}
Next, we construct a complex function $f$, whose value depends on all $A_{x_{i}}^{(i)}$:
\begin{eqnarray}
f=\left(A_{0}^{(1)}+i A_{1}^{(1)}\right)\left(A_{0}^{(2)}+i A_{1}^{(2)}\right)\cdot\cdot\cdot\left(A_{0}^{(N)}+i A_{1}^{(N)}\right).\nonumber\\
\label{aeq4}
\end{eqnarray}
Noting the properties of the sign function $v_{k}$ in Eq. (\ref{v1}), and comparing Eq. (\ref{aeq3}) with Eq. (\ref{aeq4}), we find that
\begin{eqnarray}
\sum_{\{x_{i}\}}v_{k} e^{i \sum_{i=1}^{N}{\alphazero{i}}}=Re f+Im f,
\label{aeq5}
\end{eqnarray}
where $Re f$ and $Im f$ represent the real and imaginary parts of the function $f$, respectively. Clearly, the modulus of $f$ satisfies
\begin{eqnarray}
\left | f \right |^{2}=\left(Re f\right)^{2}+\left(Im f\right)^{2}=2^{N}.
\label{aeq6}
\end{eqnarray}
Thus, we have
\begin{eqnarray}
\sum_{\{x_{i}\}}v_{k} e^{i \sum_{i=1}^{N}{\alphazero{i}}}&=&Re f+Im f\nonumber\\
&\leq&\sqrt{2\left(\left(Re f\right)^{2}+\left(Im f\right)^{2}\right)}\nonumber\\
&=&\sqrt{2^{N+1}}
\label{aeq7}
\end{eqnarray}

\vskip 0.5 cm

\emph{3. The maximum value of $\sum_{\{x_{i}\}}v_{k} e^{i \sum_{i=1}^{N}{\alphazero{i}}}$ for $N=4,5,6,7,8$.}

We used \emph{Mathematica} to calculate the maximum value of $\sum_{\{x_{i}\}}v_{k} e^{i \sum_{i=1}^{N}{\alphazero{i}}}$ when all $\alphazero{i}$'s are fixed to either $0$ or $\pi$.

(i) $N=4$

We set $\left(\alpha_{x_{4}=1}^{(4)}\right)_{0}=\pi$ and fix all other $\alphazero{i}=0$. This yields the maximum value of $\sum_{\{x_{i}\}}v_{k} e^{i \sum_{i=1}^{N}{\alphazero{i}}}$ as $4$. Substituting this into Eq. (\ref{nc4}), we calculate that for $j=1$, $\frac{\langle S_{N}\rangle_{max}}{2^{N-1}}\approx 1.10948$, which results in a violation of SI.

(ii) $N=5$

We set $\left(\alpha_{x_{5}=0}^{(5)}\right)_{0}=\left(\alpha_{x_{5}=1}^{(5)}\right)_{0}=\pi$ and fix all other $\alphazero{i}=0$. This yields the maximum value of $\sum_{\{x_{i}\}}v_{k} e^{i \sum_{i=1}^{N}{\alphazero{i}}}$ as $8$. Substituting this into Eq. (\ref{nc4}), we calculate that for $j=1$, $\frac{\langle S_{N}\rangle_{max}}{2^{N-1}}\approx 1.10948$, which leads to a violation of SI.

(iii) $N=6$

We take $\left(\alpha_{x_{6}=0}^{(6)}\right)_{0}=\left(\alpha_{x_{6}=1}^{(6)}\right)_{0}=\pi$ and set all other $\alphazero{i}=0$. This yields the maximum value of $\sum_{\{x_{i}\}}v_{k} e^{i \sum_{i=1}^{N}{\alphazero{i}}}$ as $8$. Substituting this into Eq. (\ref{nc4}), we calculate that for $j=1$, $\frac{\langle S_{N}\rangle_{max}}{2^{N-1}}\approx 1.02614$, which leads to a violation of SI.

(iv) $N=7$

We set $\left(\alpha_{x_{6}=1}^{(6)}\right)_{0}=\left(\alpha_{x_{7}=0}^{(7)}\right)_{0}=\left(\alpha_{x_{7}=1}^{(7)}\right)_{0}=\pi$ and fix all other $\alphazero{i}=0$. This gives the maximum value of $\sum_{\{x_{i}\}}v_{k} e^{i \sum_{i=1}^{N}{\alphazero{i}}}$ as $16$. Substituting this into Eq. (\ref{nc4}), we calculate that for $j=1$, $\frac{\langle S_{N}\rangle_{max}}{2^{N-1}}\approx 1.02614$, which results in a violation of SI.

(v) $N=8$

We set all $\alphazero{i}=0$, and the maximum value of $\sum_{\{x_{i}\}}v_{k} e^{i \sum_{i=1}^{N}{\alphazero{i}}}$ is obtained as $16$.
However, this result does not lead to a violation of SI, because when we substitute this result into Eq. (\ref{nc4}), we calculate that for $j=1$, $\frac{\langle S_{N}\rangle_{max}}{2^{N-1}}\approx 0.984476$.


\bibliographystyle{apsrev4-1}
\bibliography{vsisj}

\end{document}